\documentclass{aa}  
\usepackage{graphicx}
\usepackage{txfonts}
\usepackage{hyperref}  
\hypersetup{colorlinks=true,linkcolor=[rgb]{1.,0.2,0.2},citecolor=[rgb]{0.1,0.4,1.},filecolor=[rgb]{0.7,0.2,0.2},urlcolor=[rgb]{0.7,0.2,0.2}}
\usepackage{color}
\definecolor{blue}{rgb}{0., 0., 1}
\newcommand {\T}{Table\,}
\newcommand {\Sec}{Sec.\,}
\newcommand {\Fig}{Fig.\,}
\newcommand {\Eq}{Eq.\,}

\begin{document}

   \title{Detection of magnetic fields in superclusters of galaxies}

   \author{G.V. Pignataro
          \inst{1,2} \fnmsep\thanks{e-mail: \href{mailto:giada.pignataro2@unibo.it}{giada.pignataro2@unibo.it}}
          \and
          S.P. O'Sullivan\inst{4}
          \and
          A. Bonafede\inst{1,2}
          \and 
          G. Bernardi\inst{2,3,5}
          \and
          F. Vazza\inst{1,2}
          \and
          E. Carretti\inst{2}
        }

   \institute{Dipartimento di Fisica e Astronomia, Universit\`a degli Studi di Bologna, via P. Gobetti 93/2, 40129 Bologna, Italy
        \and
             INAF -- Istituto di Radioastronomia, via P. Gobetti 101, 40129 Bologna, Italy
        \and
            Department of Physics and Electronics, Rhodes University, PO Box 94, Makhanda, 6140, South Africa
        \and
            Departamento de Física de la Tierra y Astrofísica \& IPARCOS-UCM, Universidad Complutense de Madrid, 28040 Madrid, Spain
        \and
            South African Radio Astronomy Observatory (SARAO), Black River Park, 2 Fir Street, Observatory, Cape Town, 7925, South Africa}

   \date{Received September 15, 1996; accepted March 16, 1997}

 
  \abstract
   {The properties of magnetic fields in large-scale structure filaments, far beyond galaxy clusters, are still poorly known. Superclusters of galaxies are an ideal laboratory to investigate these low-density environments that are not easily identified, given the low signals large scales involved. The observed Faraday rotation measure of polarized sources along the line-of-sight of superclusters allows us to constrain the magnetic field properties in these extended environments. }
   {The aim of this work is to constrain the magnetic field intensity in low-density environments within the extent of superclusters of galaxies, using the Faraday rotation measure of polarized background sources detected at different frequencies. }
   {We selected three rich and nearby ($z<0.1$) superclusters of galaxies for which polarization observations were available at both 1.4~GHz and 144~MHz: Corona Borealis, Hercules and Leo. We compiled a catalogue of 4497 polarized background sources that either have rotation measure values from the literature, or derived from unpublished observations at 144~MHz. For each supercluster we created a 3D density cube, in order to associate a density estimate to each RM measurement. We computed the median absolute deviation (MAD) variance of the RM values grouped in three different density bins that correspond to the supercluster outskirts ($0.01<\rho/\rho_c<1$), filaments ($1<\rho/\rho_c<30$) and nodes ($30<\rho/\rho_c<1000$) regimes to investigate how variations in the RM distribution are linked to the mean density crossed by the polarized emission.}
   {We found an excess $\Delta\sigma^{2_{RRM}}_{MAD}=2.5\pm0.5$ rad$^2$ m$^{-4}$ between the lowest density regions (outside supercluster boundaries) and the low-density region inside the supercluster. This excess is attributed to the intervening medium of the filaments in the supercluster. We model the variance of the RM distribution as due to a single-scale, randomly-oriented magnetic field distribution, therefore depending upon the magnetic field intensity along the line of sight, the magnetic field reversal scale and the line-of-sight path-length. Our observations do not constrain the latter two parameters, but if we marginalized over their respective prior range, we constrain the magnetic field to be $B_{||}=19^{+50}_{-8}$ nG.}
   {Our findings are consisted with other several works which studied filaments of the large scale structure. The results suggest that the purely adiabatic compression of a primordial magnetic field, which would imply observed magnetic fields of the order of $B_{||}\sim 2$ nG, is not the only mechanism to play a role in amplifying the primordial seeds in superclusters of galaxies.}

  \maketitle
%

\section{Introduction}\label{sec:introrm}

The matter distribution of the Universe on megaparsec scales is not uniform and random, but is distributed according to a pattern called the Cosmic Web \citep{bond1996}. Galaxies and galaxy clusters are found at the high-density `nodes' of the Cosmic Web ($n_e\ll10^{-3}$ cm$^{-3}$), connected by lower-density environments such as filaments ($n_e\sim10^{-6}-10^{-4}$ cm$^{-3}$). Filaments are thought to be permeated by the warm-hot intergalactic medium \citep[WHIM;][]{gheller2015}, with temperatures of $10^5-10^7$~K, and magnetic fields. While the magnetic field properties of galaxies and galaxy clusters and groups are well studied \citep{laing2008,vanweeren2019}, the magnetic fields in the unprocessed gas in filaments and voids far beyond galaxy clusters are still poorly known. In recent years, several studies attempted to constrain the strength of magnetic fields in filaments with different approaches. Optical-IR galaxy distribution and radio cross-correlation \citep{vernstrom2017, brown2017} and image-stacking studies \citep{tanimura2020, vernstrom2021} have found estimates and limits ranging from 30 and 60 nG. With LOFAR High Band Antenna (HBA) observations, the non-detection of diffuse emission in filaments between galaxy clusters resulted in upper limit ranging from $\leq 0.2$~$\mu$G \citep{locatelli2021} to $\leq 0.75$ $\mu$G \citep{hoang2023}. Direct detection of the non-thermal synchrotron and X-ray emission associated with filaments of the cosmic web is possible \citep{vazza2019}, and it was confirmed by the discovery of two radio bridges of diffuse synchrotron emission between clusters \citep{govoni2019, botteon2020}, that resulted in an estimate of magnetic field of $\sim 0.3$~$\mu$G in the intercluster region \citep{pignataro2024b}. However, these detections represent a 'short' population of filaments ($\sim1-5$ Mpc) with overdensity $\rho/\rho_c\sim200$\footnote{$\rho_c=3H_0^2/8\pi G$ is the mean critical density of the Universe.} \citep[e.g.][]{wittor2019}, compressed by the dynamical activity between two merging clusters, while instead fainter filaments ($\rho/\rho_c\sim10$) on several tens of Mpc scale remain undetected.

An alternative approach to measure magnetic fields is to use the Faraday rotation of linearly polarized sources along the line of sight of a magnetised plasma. 
The Faraday rotation measure (RM) quantifies the rotation of the linear polarization vector as a function of wavelength and  depends on the line-of-sight magnetic field strength $B_{||}$ permeating a medium of ionised gas with electron density $n_e$, along a path length $L$ between the source at redshift $z$ and the observer:
\begin{equation}
    \text{RM}=0.812 \int \frac{n_e \, B_{||} }{(1+z)^2}\, \text{d}L     ~~~~~~[\text{rad/m}^{2}].
\end{equation}
\cite{vernstrom2023} recently reported a high polarization fraction for short-range ($<10$ Mpc) filaments in between massive halos, which implies a significantly ordered magnetic field component in these environments, consistent with the detection of an RM signal from them. \cite{carretti2024} estimated magnetic fields of $10-145$~nG with extragalactic RMs at low frequencies and overdensity $\rho/\rho_c\sim10$.
At higher frequencies, RM studies reported limits between 40~nG \citep{vernstrom2019} for extragalactic magnetic fields and 0.3~$\mu$G in superclusters of galaxies \citep{xu2006, shishir2024}. 
All these estimates and studies are of key importance in understanding the magnetogenesis scenarios in our Universe \citep{durrer2013, subramanian2016}, because the filament environments still preserve the memory of the primordial conditions, while still being more easily detectable than voids \citep{vazza2017, mtchedlidze2022}. The investigation of magnetic fields in low-density (i.e. $n_e\sim10^{-5}$ cm$^{-3}$) environments  can help distinguish between two main models: primordial scenarios, where the seed is produced during the early phases of inflation or before recombination \citep{paoletti2014}, and astrophysical scenarios, where the feedback from AGNs is responsible for the late magnetisation of the Universe \citep{bertone2006, donnert2009}.
One way to identify the location of filaments is to search for superclusters of galaxies: these are nested within the Cosmic Web, creating a coherent structure of galaxy clusters embedded in a network of filaments spanning up to hundreds of Mpcs \citep{lietzen2016, bagchi2017}. The possibility of finding filamentary structure increases within superclusters \citep{tanaka2007}, because they are characterized by higher mean densities. 

There are several different ways to identify superclusters in the sky, which makes them particularly useful when researching the large-scale structure of filaments. One frequently used approach is to exploit the galaxy overdensity distribution to trace filamentary structures connecting clusters and groups. This is supported by the vast availability of sky surveys such as the Sloan Digital Sky Survey (SDSS, \citealt{almeida2023}), the Two Micron All Sky Survey (2MASS, \citealt{skrutskie2006}), and Two Degree Field Redshift Survey (2dFRS, \citealt{huchra2012}) or the Center for Astrophysics galaxy redshift survey (CfA2, \citealt{huchra1999}).  Superclusters of galaxies are frequently identified using the Friends-of-Friends \citep[FoF, e.g.][]{zeldovich1982, einasto1984, chowmartinez2014, bagchi2017, shishir2024} algorithm, which is used to find and group points (in this case, galaxies) with unknown distributions in a simulation.  Other methods are also used, such as applying a threshold cut to the luminosity density field of the galaxy distribution \citep{einasto2007, lietzen2016}, or to the number density field constructed via Voronoi tassellation \citep{neyrinck2008, nadathur2016}. \cite{santiagobautista2020} developed an identification method based on geometrical information of the galaxy distribution.
This process of identification resulted in several different catalogues of known superclusters of galaxies that can be used in combination with Faraday Rotation measures catalogues of sources behind superclusters, to probe the magnetic field of the plasma crossed by the polarized emission of the distant radio sources \citep{xu2006, shishir2024}. 

In this work, we constrain the magnetic field intensity in filaments by using RM measurements of sources behind galaxy clusters.
In \Sec\ref{sec:dataandobs} we present the available RM grid data and the construction of the polarized source catalogue behind the line of sights of the selected superclusters; in \Sec\ref{sec:methods} we show how we constructed the supercluster density maps and how we combine RM values from different surveys; in \Sec\ref{sec:resultrrm} and \Sec\ref{sec:discussion} we present the results of the analysis of the trend of RRM variance and the gas density, and discuss different scenarios to constrain the magnetic fields in low density environments; finally, in \Sec\ref{sec:conclusions}, we summarize our findings. 

\section{Observations and dataset}\label{sec:dataandobs}

In this section, we describe the different datasets that were used in this work. 
    \subsection{Selection of superclusters of galaxies}\label{sec:mscc}

\begin{figure}
   \centering
   \includegraphics[width=1\linewidth]{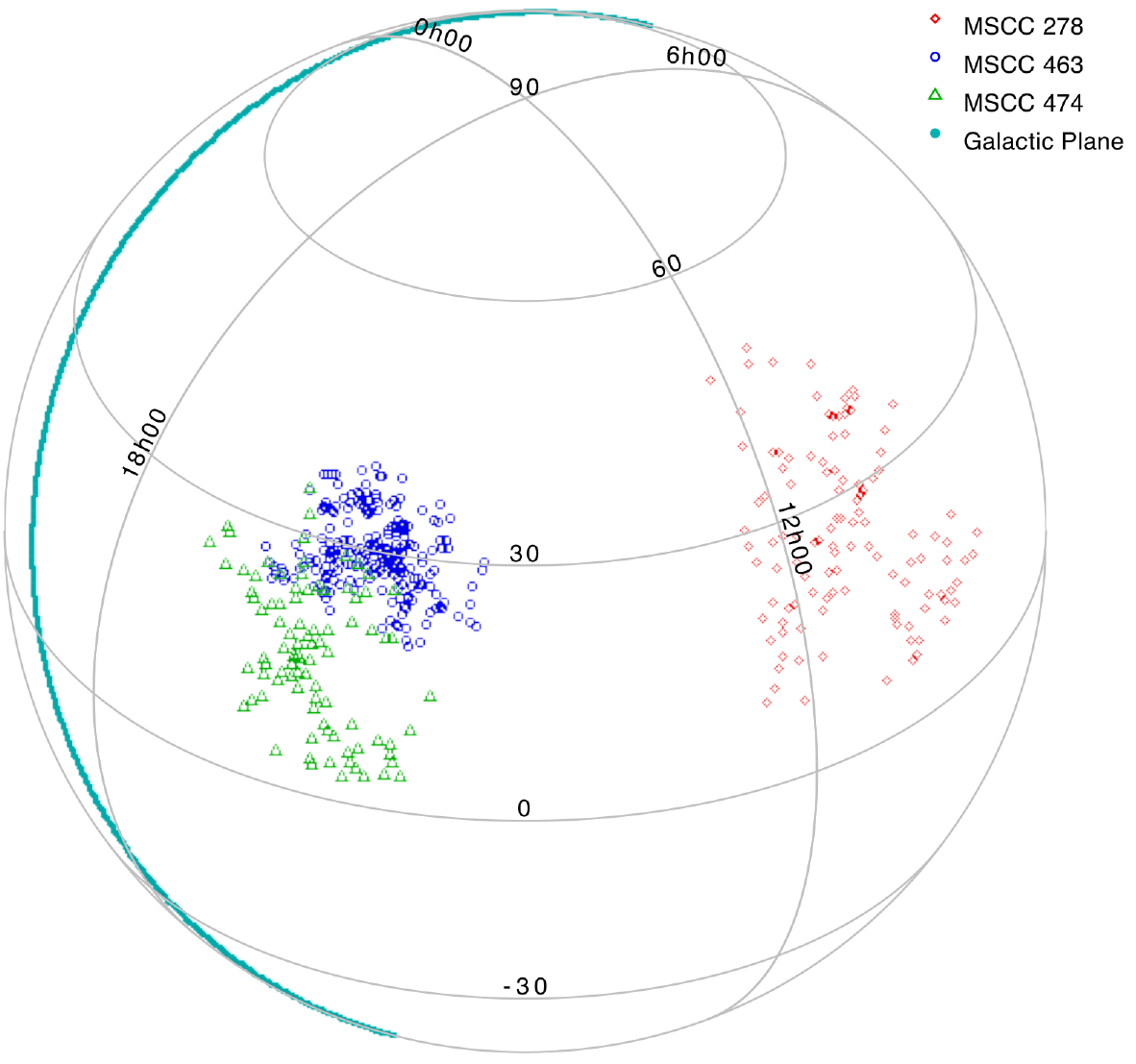}
      \caption[Sky distribution of the cluster members in each supercluster of galaxies.]{Sky distribution of the supercluster members in each supercluster in equatorial reference system. The location of the Galactic Plane is shown in light-blue.}
         \label{fig:mscc_sky}
\end{figure}

This work is based on the selection of polarized sources for which we have an RM value, in the line of sight of superclusters of galaxies. We based this study on the LOFAR Two-Metre Sky Survey (LoTSS) RM Grid \citep{osullivan2023}, which is sensitive to small RM, therefore we select nearby ($z\leq0.1$) superclusters in the Northern sky that are covered, at least partially, by LoTSS observations. 
We chose to analyse three rich superclusters: \textit{Corona Borealis}, \textit{Leo}, and \textit{Hercules}. These three superclusters are part of the all-sky Main SuperCluster Catalogue \citep[MSCC,][]{chowmartinez2014}, which is a catalogue of 601 superclusters created with the combination of a compilation of the rich Abell clusters \citep{adernach2005} and spectroscopic redshifts for galaxies in the SDSS-DR7 \citep{abazajian2009}. With a tunable Friends-of-Friends algorithm, \cite{chowmartinez2014} are able to provide a full list of the cluster members with their redshift, coordinates and supercluster membership out to a redshift of z=0.15. This catalogue was further expanded by the analysis of \cite{santiagobautista2020}, where they select 46 MSCC clusters to map the elongated structures of low relative density inside each supercluster and they employ optical galaxies with spectroscopic redshifts from the SDSS-DR13 \citep{albareti2017}. The SDSS galaxies are selected inside a volume of a box with 'walls' set to a distance of 20 $h^{-1}_{70}$ Mpc from the center of the farthest clusters in each direction, for each supercluster \citep{santiagobautista2020}.
Therefore, with this catalogue we are able to map the galaxy density over the supercluster volume as well as to recover the redshift, virial radius, and location of each supercluster member (clusters and groups of galaxies). 
The properties of the selected superclusters, as reported in 
\cite{santiagobautista2020}, are summarized in \T\ref{tab:mscc}, and the nominal location of each supercluster member on the sky is shown in \Fig\ref{fig:mscc_sky}.

\begin{table*}[]
\renewcommand{\arraystretch}{1.1}

\centering
\resizebox{\textwidth}{!}{%
\begin{tabular}{cccccccc}
{Name} & {MSCC-ID} & {RA, Dec} & \textbf{$<z>$} & \textbf{$N_{mem}$} & \textbf{$N_{Cl}$} & \textbf{$V_{box}$} & \textbf{$V_{fil}/V_{box}$} \\
&& {[deg,deg]} &&&&[$10^3$ $h^{-3}_{70}$ Mpc$^3$]&[\%] \\ \hline \hline
\textit{Corona Borealis} & 463              & 232.18, 30.42                                                             & 0.073          & 226                & 14                & 959.2                                                                                           & 1.1                                                                           \\
\textit{Leo}             & 278              & 169.37, 28.34                                                             & 0.033          & 115                & 6                 & 459.3                                                                                           & 1.6                                                                           \\
\textit{Hercules}        & 474              & 241.56, 16.22                                                             & 0.036          & 90                 & 5                 & 343.8                                                                                           & 0.9                                                                           \\ \hline
\end{tabular}%
}
\smallskip
\caption[MSCC superclusters properties]{MSCC superclusters properties. We list the superclusters selected for this study (1) names, (2) their MSCC-ID \citep{chowmartinez2014} , (3) their coordinates (J2000), (4) the mean redshift, (5) the total number of members, (6) the number of Abell/ACO clusters, (7) the volume of the box containing the SDSS galaxies, and (8) the filling factor of filaments as found in  \cite{santiagobautista2020}. }
\label{tab:mscc}
\end{table*}

    \subsection{RM Grids }\label{sec:rmgrids}

Faraday rotation measure grids are a valuable tool to study the origin and evolution of cosmic magnetism, in particular by measuring the properties of extragalactic magnetic fields \citep{vernstrom2019, osullivan2019}. In this particular framework, we observe linearly polarized radio sources across a particular area of the sky, covered by the extension of a nearby supercluster, to investigate the Faraday rotation properties of the low-density, large-scale environments.
To do this, we are mainly interested in the RM variance generated by the variations of the Faraday rotating medium along different lines of sight, either due to the intergalactic or local media. To isolate this component, it is necessary to remove the Galactic contribution from the Milky Way, and the errors introduced by the measurements. 
The RM variance has been investigated extensively with the 37543 RM values from the NRAO VLA Sky Survey data \citep[NVSS,][]{condon1998,taylor2009} at 1.4~GHz, to characterize both the Milky-Way properties \citep[e.g.][]{purcell2015, hutschenreuter2019}, and to isolate the RM variance contribution local to the source itself \citep{rudnick2003, osullivan2013, anderson2018, banfield2019, knuettel2019}.

RM studies at metre wavelengths offer a significant advantage over centimetre wavelength observations due to a better RM accuracy and sensitivity to low RM values. The accuracy of Faraday rotation measurements is directly related to the wavelength-squared coverage, thus RM studies at metre wavelengths provide substantially higher accuracy for individual RM measurements \citep{neld2018, osullivan2018, vaneck2018}. Despite this, identifying linearly polarized sources at long wavelengths presents its own challenges, primarily requiring high angular resolution and high sensitivity to counteract the significant effects of Faraday depolarization which results in a smaller fraction of radio sources exhibiting detectable polarization levels \citep[e.g.,][]{farnsworth2011,bernardi13,lenc2017}. LOFAR is addressing these challenges effectively with its capability to produce high-fidelity images at high angular resolution \citep{morabito2016, jackson2016, harris2019, sweijen2022}. Additionally, the LOFAR wide field of view and large instantaneous bandwidth facilitate the efficient surveying of large sky areas, aiding the detection of numerous linearly polarized sources and their RM values. The LoTSS-DR2 RM Grid \citep{osullivan2013} has produced a catalogue of 2461 extragalactic high-precision RM values.
The integration of RM Grid catalogues from both metre and centimetre wavelengths is crucial for a more comprehensive understanding of the various contributors to Faraday rotation along the line of sight. We are interested in this combined approach, and we use of both the NVSS and LoTSS RM Grids to identify sources in the line of sight of our selected superclusters. The three superclusters have different extensions on the sky. We select all sources inside a circle of radius $19^\circ$ centered on each supercluster position.
We check for duplicates between NVSS and LoTSS RM sources, and also take into account the slight overlap between Corona Borealis and Hercules on the plane of the sky, which could introduce a number of additional duplicates. The duplicated RM were removed from the NVSS catalogue and kept in LoTSS catalogue, leaving us with 3679 RM values from the NVSS and 579 RM values from the LoTSS DR2 Grid. 
For the 3679 NVSS RMs, we recompute their overestimated errors $\sigma^{NVSS}_{RM}$ reported in \cite{taylor2009}, following the equation of \cite{vernstrom2019} 
\begin{equation}
    \sigma^{NVSS}_{RM}=150 \, \frac{\sqrt{2}\sigma_P}{P} \, \, \text{rad m}^{-2},
\end{equation}
where $P$ is the polarized intensity\footnote{The polarized intensity is defined as $P=\sqrt{Q^2+U^2}$, where $Q$ and $U$ are the Stokes parameters.} of the source and $\sigma_P$ the associated error.

\subsection{Source selection from non-DR2 fields}\label{sec:nondr2}
    
The selected superclusters are very extended across the sky, in particular Hercules (ID 474) and Leo (ID 278) can reach a Declination as low as $+3$° (see \Fig\ref{fig:mscc_sky}). The LoTSS-DR2 sky area imaged in polarization covers 5720~deg$^2$ and it is split between two fields centred at $0^{\rm h}$ and $13^{\rm h}$ respectively, down to a Declination of approximately $+15$° \citep{shimwell2019, osullivan2023}. Therefore, part of these superclusters are not covered by the DR2. To improve the statistics and to find additional linearly polarized sources to probe the supercluster environments, we analysed an additional 177 LoTSS pointings not included the publicly available DR2 sky area. These RMs will form part of the LoTSS DR3 RM Grid data release.
The analysis of the additional pointings follow the same procedure used to create the LoTSS-DR2 RM Grid, as reported in \cite{osullivan2023}. Here, we report the main steps:
\begin{itemize}
    \item We used Stokes Q and U image cubes at $20''$ resolution and the Stokes I $20''$ resolution images and source catalogues \citep{williams2019,shimwell2022} from the LoTSS initial data products \citep{shimwell2019, tasse2021}; 
    \item The RM synthesis technique \citep{burn1996, brentjens2005} was applied on the Q and U images using \texttt{RMsynt1D} from \texttt{RM-Tools}\footnote{\href{https://github.com/CIRADA-Tools/RM-Tools}{https://github.com/CIRADA-Tools/RM-Tools}} \citep{purcell2020} with uniform weighting, for pixels in the Stokes I $20''$ image where the total intensity was greater than 1~mJy~beam$^{-1}$. 
    Initially, it is necessary to define a 'leakage' exclusion range between $\pm3$~rad~m$^{-2}$ to remove the contamination of the instrumental polarization in the Faraday depth spectrum or Faraday dispersion function (FDF). The leakage peak occurs intrinsically at $0$ rad m$^{-2}$ with a degree of polarization of $<\sim 1\%$ of the Stokes I intensity \citep{shimwell2022}, but it is shifted by the typical LOFAR ionospheric RM correction up to 3~rad~m$^{-2}$ \citep{sotomayor2013, snidaric2023}.
    To help reduce the effect of sidelobe structure, we run the code \texttt{RMclean1D} \citep{purcell2020}, that performs deconvolution of the FDF. With a variation of the H\"ogbom \texttt{CLEAN} \citep{hogbom1974} algorithm, the \texttt{RM-clean} iteratively subtracts scaled versions of the rotation measure spread function (RMSF) from the FDF until a noise threshold is reached (\citealt{heald2009}, for a full description). We used a threshold of four times the noise in the FDF during the \texttt{RM-clean} process. 
    From the clean spectrum and outside the leakage range, we identify the peak polarized intensity for each pixel in the output cube of the FDF. 
    \item We estimated the noise $\sigma_{QU}$ and initially consider the Faraday depth (i.e. the RM) value corresponding the polarized intensity peaks in the FDF larger than $5.5\sigma_{QU}$ from the rms of the wings of the real and imaginary parts of the FDF ($\phi< -100 \text{ rad m}^{-2}$ and $  \phi > 100 \text{ rad m}^{-2}$). Finally, an RM image, a polarized intensity image and a degree of polarization map were created. From the polarized intensity image, all pixels within a box of $20\times20$ pixels and above the selected noise threshold, were grouped together. The highest signal to noise pixel in this group is the catalogued RM value and sky position of this source component.
\end{itemize}
An initial review of the catalogued sources revealed that many bright sources detected at small Faraday depths with very low degree of polarization were likely instrumental peaks extending beyond the previously excluded leakage range. Consequently, to eliminate many of these sources we extended the leakage range to $-5$ rad m$^{-2} < \phi < +5$ rad m$^{-2}$ and excluded those sources with fractional polarization smaller $p<2$\%, or very high values $p>30$\%.
To be conservative, an $8\sigma_{QU}$ threshold was then applied following what is expected from false detection rates, which is as low as $10^{-4}$ at $8\sigma_{QU}$, against a possible 4\% at $5\sigma_{QU}$ \citep{george2012}. 
After these additional cuts, we inspected the Faraday spectra of the remaining sources. From the visual inspection of the FDF and the signal-to-noise ratio (SNR) it is possible to find some unreliable RM values that fell out the cuts. In particular, we inspect the FDF of sources very close-to-the-acceptance threshold of $8\sigma_{QU}$, and check also the degree of polarization. An example of accepted and rejected source after the visual inspection of the Faraday spectrum is shown in \Fig\ref{fig:plots_insp}. In this case, the rejected source satisfies the $8\sigma_{QU}$ criterium but the FDF shows the presence of several other peaks at a similar SNR level, making the detection unreliable.

The preliminary catalogue was then revised, and we noticed that some fields contained a high number of sources with very similar RM values and low polarization fraction. This is likely to be attributed to a '\textit{transfer}' of polarized flux from a bright ($>10$ mJy beam$^{-1}$) polarized source in the field, as it was noticed in the original compilation of the LoTSS-DR2 RM Grid \citep{osullivan2023}. We therefore checked each field for this effect, and removed a total of 876 unreliable candidates from all fields.

After this step, we inspected also the maps produced after running the RM synthesis, and for each source we produced cut-outs to compare the RM map, the polarized intensity image, the degree of polarization map together with the Faraday spectrum and the NVSS 45$''$-resolution Stokes I and polarized intensity contours (see \Fig~\ref{fig:maps_insp}). In this last step of inspection, we excluded sources with very complex Faraday spectrum where the peak is not clearly identified and/or is located at a pixel clearly outside the source, as well as sources with highly inconsistent LoTSS-NVSS degree of polarization. This procedure excluded a few more sources. 
Finally, we removed duplicate sources from the NVSS, as done for the LoTSS-DR2 fields, keeping the RM value with higher SNR for LoTSS-LoTSS duplicates and LoTSS RM value for LoTSS-NVSS duplicates. The final non-DR2 catalogue contains 239 polarized source components, that are added to the analysis. 
Averaging together the three superclusters fields, we have a final catalogue\footnote{The catalogue will be made available through Vizier.} with 3679 RM values from NVSS and 818 from LoTSS, for a total of 4497 polarized background sources.

\section{Methods}\label{sec:methods}
    \subsection{Density maps}\label{sec:densitymaps}
We used the final catalogue of polarized sources to study the supercluster medium by measuring the RM variance.

The RM variance is dependent on the free electron density, the line-of-sight magnetic field strength, and the  magnetic field reversals that happen along the length of the path crossed by the radiation. Therefore, it is useful to relate the measured RM values to the density of the medium in the supercluster, and investigate which combinations of parameters will yield the observed variance. 
To limit the supercluster extent and estimate the density in each region, we have used the gas density profile of each supercluster member.
For this work, we chose to use the Universal gas density profile for galaxy clusters presented in \cite{pratt2022}. 
Using XMM-Newton observations, they derive an average intracluster medium density profile for a sample of 93 Sunyaev-Zeldovich effect -selected systems and determine its scaling with mass and redshift.
The median radial profile is a function of a scaled radius $x=R/R_{500}$, and is expressed as the product between a normalisation that can vary with the redshift $z$ and the mass $M_{500}$ (i.e. the total mass within the radius $R_{500}$, i.e. when the mean matter density is $500$ times the critical density of the Universe):
\begin{equation}
    \rho_{m}(x,z,M_{500})=N(z,M{500}) \, f(x),
\end{equation}
where $f(x)$ has the shape of a generalised Navarro-Frenk-White (GNFW) model \citep{nagai2007}:
\begin{equation}
    f(x) = \frac{f_0}{\left( \frac{x}{x_s} \right)^{\alpha} \left[ 1 + \left( \frac{x}{x_s} \right)^{\gamma} \right]^{\frac{3\beta - \alpha}{\gamma}}}
\end{equation}
where $x_s$ is the scaling radius, and the parameters $\alpha$, $\beta$, $\gamma$ are the slopes at  $x \ll x_s$, at $x \gg x_s$, and at $x\sim x_s$, respectively. The normalisation is given by the product of $f_0$ and 
\begin{equation}
    N(z,M_{500})= E(z)^{\alpha_z}\left[\frac{M_{500}}{5\times10^{14}M_{\odot}}\right]^{\alpha_M},
\end{equation}
where $E(z)$ is the evolution of the Hubble parameter with redshift in a flat cosmology. 
A complete description of the fitting procedure to the 93 SZ-selected system can be found in \cite{pratt2022}, while here we only report the best-fit model parameters: \\
\begin{equation}
\begin{aligned}
f_0 &= 1.20 \pm 0.15 \text{,} \\
x_s &= 0.28 \pm 0.01 \text{,} \\ 
\alpha &= 0.42 \pm 0.06 \text{,} \\ 
\beta &= 0.78 \pm 0.03 \text{,} \\ 
\gamma &= 1.52 \pm 0.16 \text{,} \\
\alpha_z &= 2.09 \pm 0.02 \text{, and} \\
\alpha_M &= 0.22 \pm 0.01 \text{.} 
\end{aligned}
\end{equation}
The deprojected density profile with the best-fit parameters is applied to each supercluster member and computed radially for each element in a 3D cube of axis (RA, Dec and $z$). We zero-padded the density profile at distances $\geq10$ $R_{500}$ from each cluster center.
With this approximation, we are treating all members as galaxy clusters, including small groups of galaxies down to five members; in this way, while the contribution of the smallest systems will be of low impact on the density at large distances from their centers, it helps trace the large-scale structure of the supercluster. We are not implementing an additional density component for the filaments between clusters (see \Sec\ref{sec:discussion}). 

\begin{figure*}[h!]
   \centering
   \includegraphics[height=0.95\textheight]{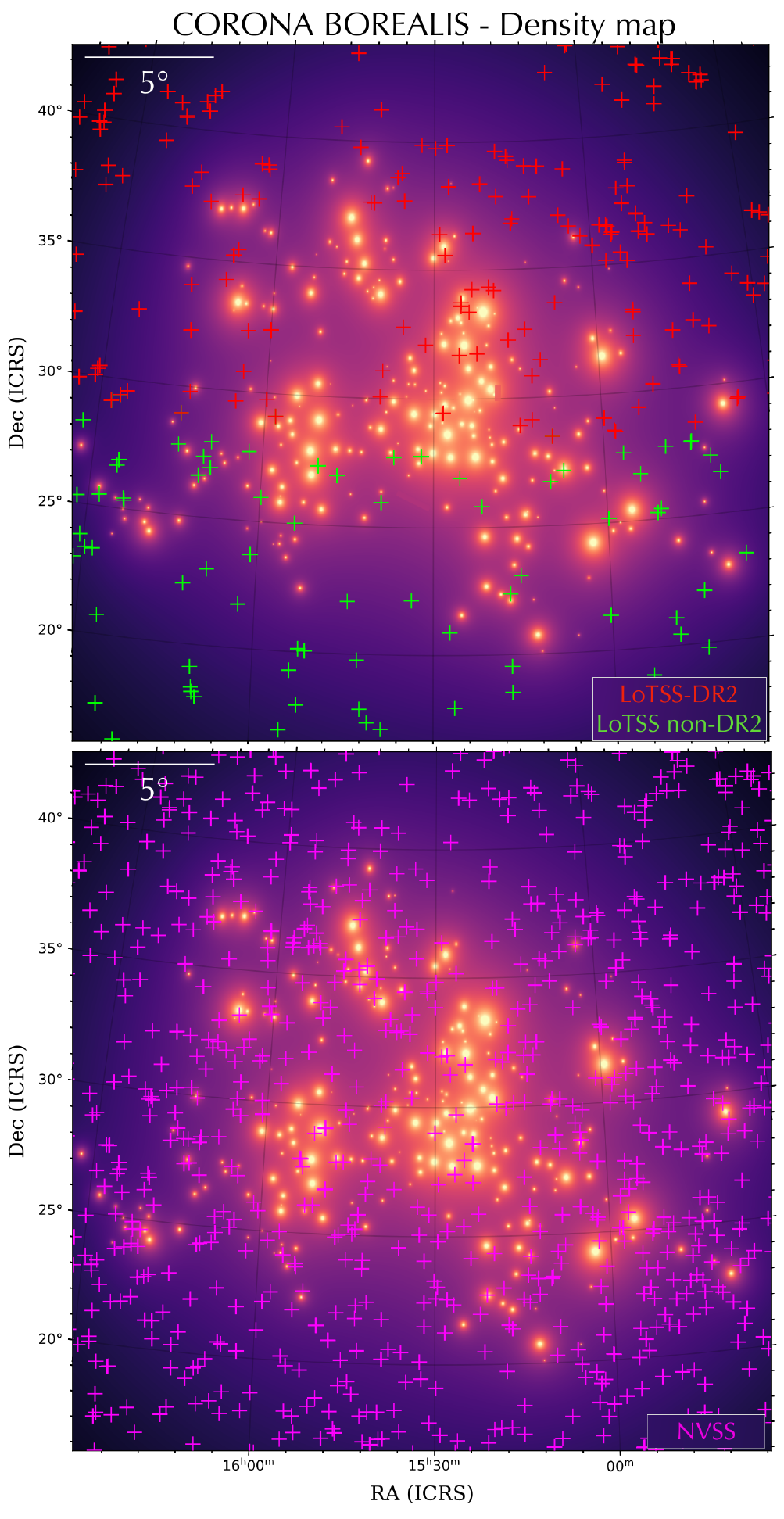}
   \caption[Two-dimensional density map for the Corona Borealis supercluster]{Two-dimensional density map for Corona Borealis supercluster, assuming that each member is on the same redshift plane. The location of the RM sources is shown in red for LoTSS-DR2, green for LoTSS non-DR2 and magenta for NVSS.}
         \label{fig:cbor_map}
\end{figure*}

\begin{figure*}[h!]
   \centering
   \includegraphics[height=0.95\textheight]{ 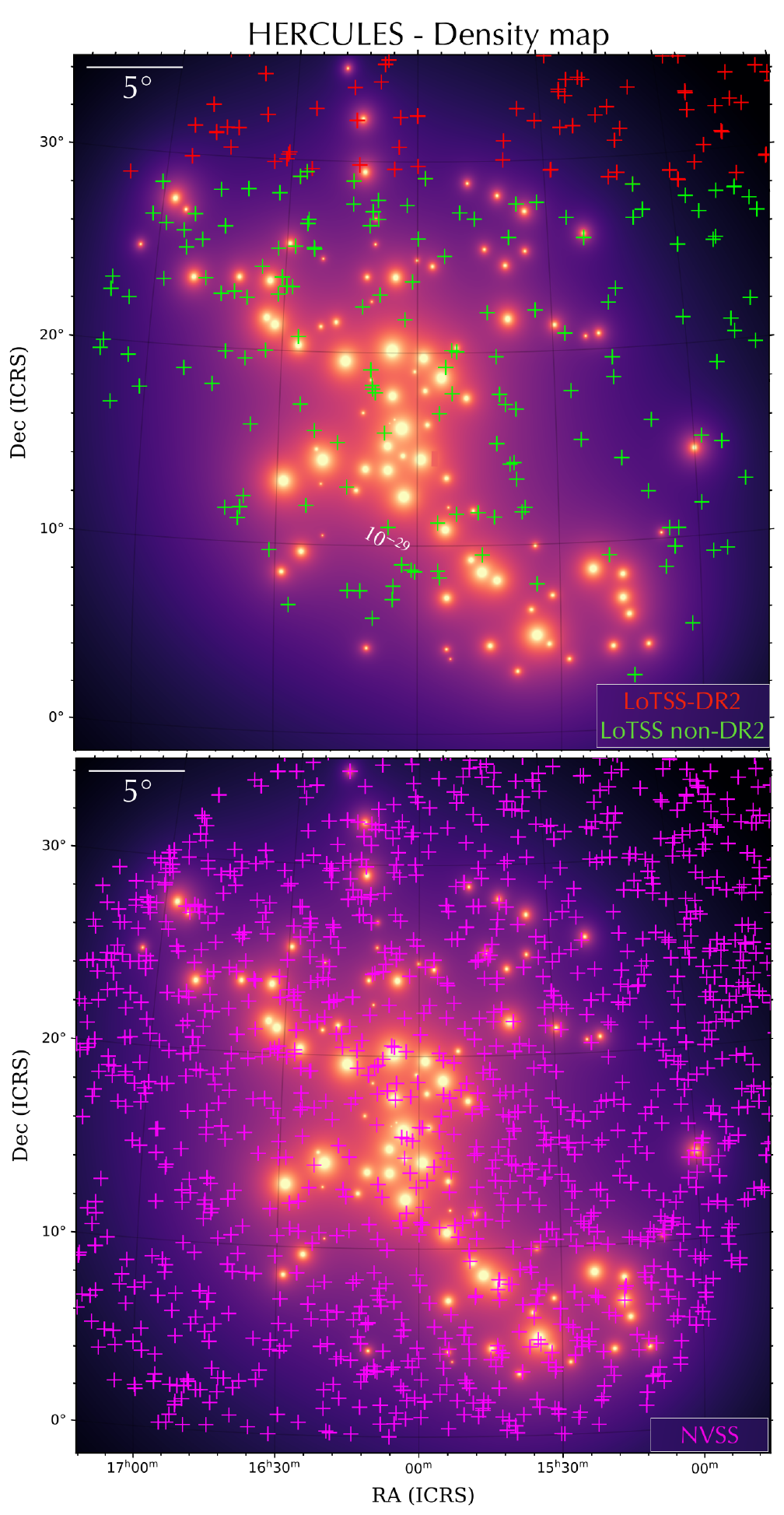}
   \caption[Two-dimensional density map for Hercules supercluster]{Same as \Fig\ref{fig:cbor_map}, but for the Hercules supercluster.}
         \label{fig:her_map}
\end{figure*}

\begin{figure*}[h!]
   \centering
   \includegraphics[height=0.9\textheight]{ 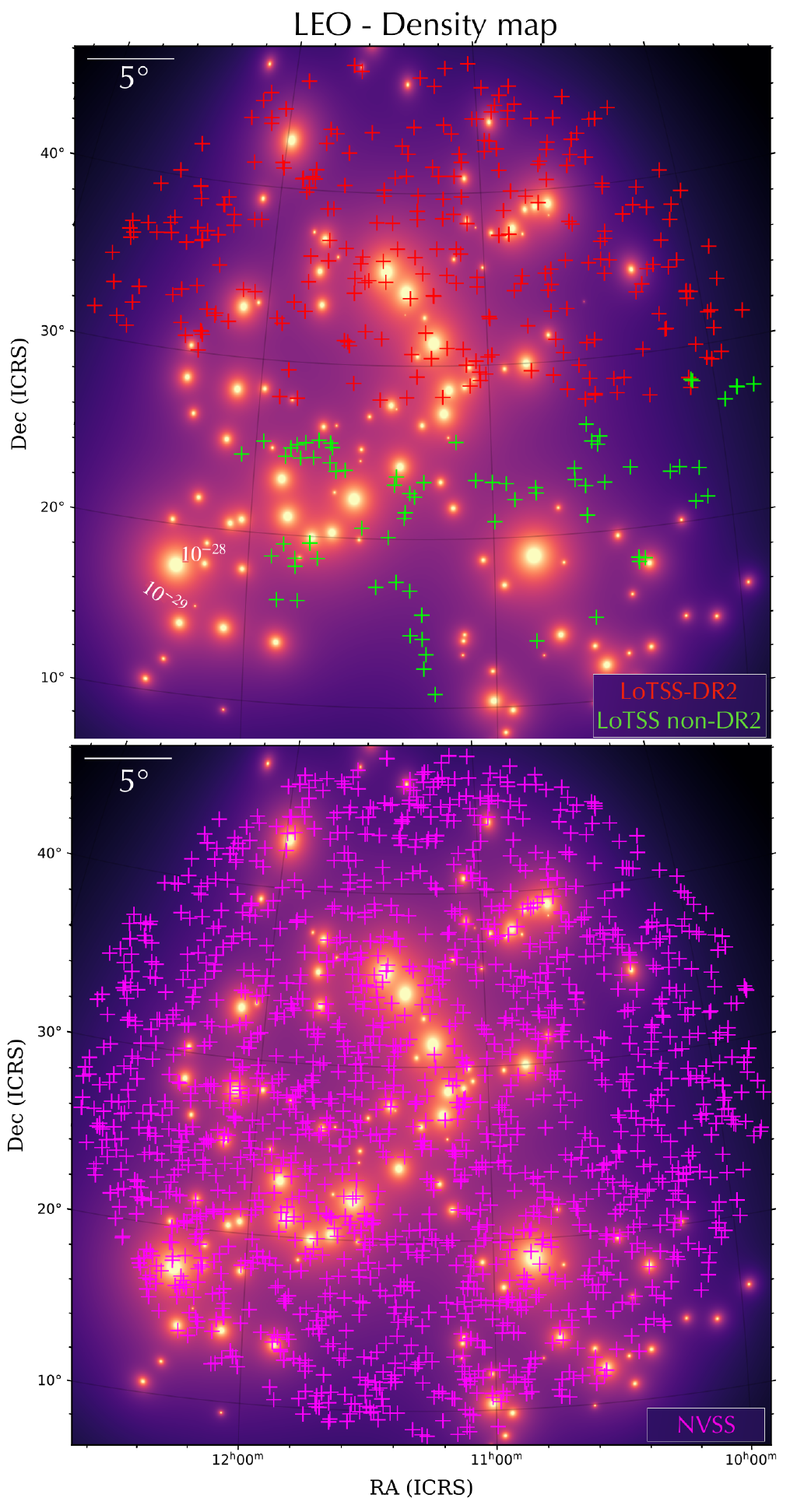}
   \caption[Two-dimensional density map for Leo supercluster]{Same as \Fig\ref{fig:cbor_map}, but for the Leo supercluster.}
         \label{fig:leo_map}
\end{figure*}

The construction of the density cubes for each supercluster allows us to describe the boundaries of the supercluster and investigate whether the radiation from the selected polarized background sources is crossing regions of high or low mean density, inside or outside the supercluster. Therefore, we computed a mean density $\overline{\rho}$ along the direction of each radio source, averaged over the redshift interval of each cube, that allows us to bin the sources in different density regimes. A two-dimensional representation of the density maps for each supercluster is shown in \Fig\ref{fig:cbor_map}, \Fig\ref{fig:her_map}, and \Fig\ref{fig:leo_map}.
It is noticeable how the sources found at 144~MHz have a smaller areal number density ($0.43$ deg$^{-2}$, \citealt{osullivan2023}) with respect to the ones found at 1.4~GHz ($>1$ deg$^{-2}$, \citealt{taylor2009}). At both frequencies, sources are rarer in the highest density regions, with $\sim10\%$ and $\sim13\%$ of the total number of sources found at mean densities $>\sim10^{-27.5}$ g cm$^{-3}$, at 144 MHz and 1.4 GHz respectively (see \Sec\ref{sec:resultrrm}). 
This is to be expected, due to the strong effect of Faraday depolarization in galaxy clusters that is more important at lower frequencies \citep{osullivan2019, carretti2022}. 

With this selection, we are also probing the regions outside supercluster boundaries; the sources in the line-of-sight of very low density environments serve as a control sample that allows us to quantify how the supercluster structure is contributing to the RM variance of the higher density regions sources.

\subsection{Statistics of the RM population}\label{sec:weighting}

The accuracy with which we can isolate the effects of the supercluster structure on the RMs of the background sources depends on the size of our sample and the dispersion of the RM distribution. 
The measured RM is the combination of the Galactic RM (GRM) component, the extragalactic component (RM$_{ext}$) and a noise term. We are mainly interested in the extragalactic component, that can be either attributed to the medium local to the source \citep{laing2008}, or to the foreground intergalactic medium. Therefore, we subtract off the GRM component to be left with a residual RM (RRM):
\begin{equation}
    \text{RRM} = \text{RM} - \text{GRM},
\end{equation}
where the GRM is estimated as the median of a disc of radius of $0.5$° centred at the source position from the GRM map by \cite{hutschenreuter2022} from several extragalactic source RM catalogues, including LoTSS and NVSS. Following \cite{carretti2022}, the choice of the radius is approximately the average spacing between sources in the catalogue suite used by \cite{hutschenreuter2022}. The median over a region can mitigate the possible effects of using sources that are part of the catalogue suite used to infer the GRM map, that results in the GRM at the exact source position to be slightly biased towards the source RM \citep{carretti2022}. While some residual contributions from GRM variations on smaller angular scales are possible, these are already mitigated by the large number of sources in our dataset, which cover different and extended regions of the sky, well away from the Galactic Plane (\Fig\ref{fig:mscc_sky}). Therefore, we are confident that by averaging over these sources, the potential scatter introduced by correlations does not necessarily introduce a systematic bias on the results.\par 
We can estimate the RRM spread of the population, for example, with a median absolute deviation (MAD) statistic, which is less sensitive to outliers in a distribution. The MAD can be used analogously as the standard deviation (see, e.g., \citealt{stuardi2021}) to measure the dispersion of the distribution, by introducing a constant scale factor:
\begin{equation}
    \sigma_{MAD} = k \cdot \text{MAD} \approx 1.48 \cdot \text{MAD}
\end{equation}
where the value of $k$ is taken assuming normally distributed data.
Therefore, the intrinsic RRM MAD variance is obtained by subtracting the squared total noise term, consisting of the measurement term ($\sigma^2_{RM, err}$) and the GRM error ($\sigma^2_{GRM, err}$), from the observed RRM variance: 
\begin{equation}
    \sigma_{MAD}^{2_{RRM}} = \sigma_{MAD}^{2_{RRM, obs}} - \sigma^2_{RM, err} -\sigma^2_{GRM, err} .
\end{equation}
As an indication, the typical measurement error for the RM in the selected LoTSS sources is $\sim0.06$ rad m$^{-2}$, as opposed to $\sim10$ rad m$^{-2}$ for the NVSS sources, while the average GRM error is $\sim0.7$ rad m$^{-2}$.
The combination of the LoTSS RM Grid and NVSS RM catalogue can achieve a better sensitivity for the purpose of investigating the magnetic field strength and structure in superclusters. However, the different variances of the populations, for example, due to different survey sensitivities, frequency, foreground screens and background source properties must be properly weighted for them to be combined in the most effective manner \citep{rudnick2019}. In particular, we aim to give more weight to the sources that are better probes of the foreground medium in the supercluster, while minimizing the local RM variations. One useful proxy of small RM scatter is the fractional polarization of a radio source, which is dependent on both the intrinsic degree of order of the magnetic field and the Faraday depth structure across the emission region. Additionally, the presence of fluctuations on the small scale can cause Faraday depolarization from the mixing of different polarization vector orientations within the beam, which will reduce the fractional polarization. 
The RM variance is directly linked to the Faraday depolarization, as expressed in \citep{burn1996} for an external screen:
\begin{equation}
    p(\lambda)=p(\lambda=0)e^{-2\lambda^4\sigma^2_{RM}}.
\end{equation}
This will result in a correlation between fractional polarization and depolarization \citep[e.g.][]{stuardi2020}. Higher fractional polarization implies smaller RM variations due to the medium local to the source, and also smaller scatter in the RM distribution \citep{lamee2016}. LoTSS detections are already preferentially selecting low depolarization sources with minimum Faraday complexity  \citep{osullivan2023}, which are excellent probes for our aim. However, the NVSS RM sources show more complexity and different polarization structure due to the nature of the host galaxy \citep{osullivan2017}. To reduce the effect of low fractional polarization sources on the extragalactic RM variance, we can separate the NVSS sources into two populations, based on the median degree of polarization of the sources in the NVSS RM catalogue ($\sim5$\%). Therefore, we will weigh differently three populations: LoTSS sources (population \textit{a}), NVSS sources with high degree of polarization (population \textit{b}, $p>5$\%) and NVSS sources with low degree of polarization (population \textit{c}, $p<5$\%).
Following \cite{rudnick2019}, each population can be described with its own intrinsic RRM variance $\sigma^2_{i}$ and a corresponding uncertainty, $\delta_i$, calculated as:
\begin{equation}
    \delta_i=\sqrt{\frac{2}{N_i}}\sigma^2_{i},
\end{equation}
where $N_i$ is the number of sources of each population.
We obtained the variance of the whole sample as the inverse-variance-weighted average of the variance of the three populations:
\begin{equation}
\sigma^2_{MAD, tot}(RRM)=\frac{\left(\frac{\sigma^2_{MAD,a}}{\delta^2_a} + \frac{\sigma^2_{MAD,b}}{\delta^2_b} + \frac{\sigma^2_{MAD,c}}{\delta^2_c} \right)}{ \left( \frac{1}{\delta^2_a}+ \frac{1}{\delta^2_b}+ \frac{1}{\delta^2_c}\right)},
\end{equation}
and the total uncertainty
\begin{equation}
    \delta_{tot}=\frac{1}{\sqrt{\left( \frac{1}{\delta^2_a}+ \frac{1}{\delta^2_b}+ \frac{1}{\delta^2_c}\right)}}.
\end{equation}
Under this choice, the LoTSS sample will be weighted more than the NVSS sample.

\section{Results and discussion}\label{sec:discussion}
    \subsection{RRM variance vs. density}\label{sec:resultrrm}
We want to investigate the trend between the variations in the RRM distribution of the populations of sources through the line-of-sight of superclusters of galaxies, and the density of the medium crossed by the polarized emission. As explained in \Sec\ref{sec:densitymaps} and \Sec\ref{sec:weighting}, with the construction of the 3D  density  we are able to associate a mean value of density to each source in the field of the each supercluster, and we combine the different populations through a weighted average, where we specifically down-weight the sources we expect to have a large RM variance in their local environments. Therefore, we can maximize the resulting accuracy of any possible RM signature from the superclusters by binning all the sources from the three superclusters in three main density regimes, thereby improving the MAD statistics. 

The task of identifying and describing the cosmic web components has been undertaken through numerical simulations and observations with several different methods, both investigating the global pattern in a statistical way \citep[see e.g.][]{peacock1999, hoyle2002, colberg2007}, and segmenting the structure into its morphological components: voids, filaments, and clusters \citep[e.g.][]{stoica2007,stoica2010, genovese2010, gonzalez2010, cautun2013}. We can base our investigation of supercluster density regimes on the density distribution across cosmic environments presented in \cite{cautun2014}, where they show that various structures are characterized by different density values. The node regions, where clusters reside, are typically of the highest density ($\rho/\rho_c \geq 100$) and filaments also represent over-dense regions spanning a wide range of density values ($\rho/\rho_c \sim 1-10$), while voids are very under-dense ($\rho/\rho_c \leq 1/10$).
Although it is difficult to precisely define a density range to describe exactly the supercluster environments, we use these density regimes to define the bin ranges for our sources. Specifically, the RRMs that have lines of sight inside the virial radii of the galaxy clusters (nodes) and those with line of sight through the low-density gas inside the supercluster boundaries but outside the virial radii of clusters (filaments). We compare with the sources that fall outside the supercluster boundaries, assuming their RRMs can be attributed to typical extragalactic background. We can use their RRMs to identify the contribution of the supercluster and investigate the magnetic fields in these regions. 
\begin{figure}[h!]
    \centering
    \includegraphics[width=1\linewidth]{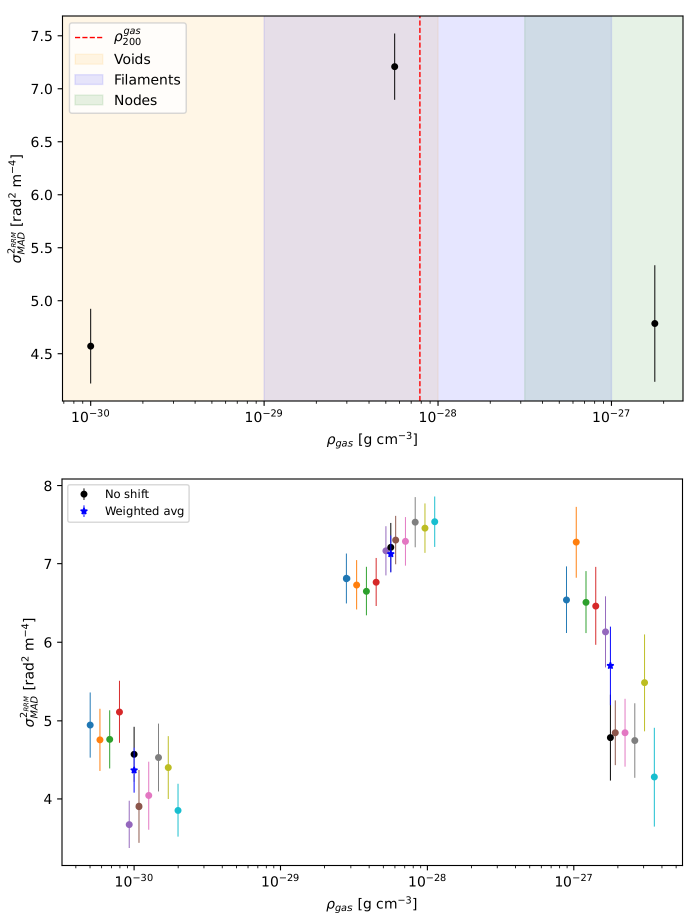}
    \caption[\textit{Top panel}: Total weighted RRM MAD variance trend with gas density in superclusters of galaxies]{Total weighted RRM MAD variance $\left(\sigma^{2_{RRM}}_{MAD}\right)$ trend with gas density in superclusters of galaxies. The plot shows the resulting variance from binning the sources in different gas density regimes. The background is divided between the ranges of densities that are typically related to voids (yellow), filaments (blue), and nodes (green) \citep{cautun2014}. The red dashed line represents where the gas $ \rho_{200}$ limit would approximately be, to highlight the density trend outside galaxy clusters. We measure an excess in the RRM variance between the first and second density bins, that can be attributed to the contribution of the low-density magnetised gas in the supercluster structure. \textit{Bottom panel}: Same as top panel, but varying the bins edges with steps of 0.03 (-0.03) with respect the original chosen value. The different resulting MAD variances and their uncertainties are show in different colors. The weighted average of these results in each bin is shown in blue, consistent with the original value shown in black. }
    \label{fig:totalsigmaRRM}
\end{figure}

We define three starting bins based on the previous considerations: $10^{-31}< \rho_{gas} < 10^{-29}$ g cm$^{-3}$ ($0.01<\rho/\rho_c<1$) for sources that can be considered outside supercluster boundaries, $10^{-29}< \rho_{gas} < 10^{-27.5}$ g cm$^{-3}$ ($1<\rho/\rho_c<30$) for sources that are falling inside supercluster boundaries but outside galaxy clusters, and finally $10^{-27.5}< \rho_{gas} < 10^{-26}$ g cm$^{-3}$ ($30<\rho/\rho_c<1000$) for sources that are inside galaxy clusters virial radii. 
For each bin, we compute the weighted MAD standard deviation of RRMs from LoTSS, NVSS high-degree of polarization, and NVSS low-degree of polarization source populations (see \Sec\ref{sec:weighting}). We show the resulting total weighted MAD variance $\left(\sigma^{2_{RRM}}_{MAD}\right)$ as a function of the gas density in \Fig\ref{fig:totalsigmaRRM} (top panel). If we consider the RRMs variations of sources falling outside supercluster boundaries (first bin) to be mostly caused by the medium local to the source, then we can subtract the local contribution to the variance in the second bin to isolate the effect of the supercluster low density regions. Between the first and second bin, we measure an excess RRM variance of $\Delta\sigma^{2_{RRM}}_{MAD}=2.5\pm0.5$ rad$^2$ m$^{-4}$. \par
To investigate the effect of the choice of bin boundaries in this detection, we can vary the bin edges to shift the boundaries of different quantities and see the effect on the resulting variance. We choose to introduce a shift between 0.0 and 0.3 (-0.3) in 10 equal logarithmic step. The result of this test are shown in \Fig\ref{fig:totalsigmaRRM} (bottom panel), where the results of each shift are marked in a different color. The bin that is most affected by the choice of bin edges is the highest density one, that exhibit very different results. It can be noted that moving the bin edge towards higher density values (i.e. toward galaxy cluster centers), the variance gets lower. This is consistent with the presence of only few sources near cluster centers with low variance that survive depolarization effects. In general, this bin shows higher uncertainty due to the smaller number of sources (only 564 RMs, while 1195 and 2738 RMs in the first two bins) found at higher densities for depolarization effects (see e.g. \citealt{bonafede2011, boringer2016, osinga2022}), and therefore we will not be considering these regions for the purpose of this analysis. If we compute the weighted average of the resulting $\sigma_{MAD}$ and their uncertainties, we find that the first two bin results are in agreement with the starting value.

    \subsection{Constraints on supercluster magnetic fields}

With the detection of an excess in RRM variance that can be attributed to the supercluster structure, we attempt to estimate the required field strength permeating the low-density environments crossed by the polarized emission.
The variance of the RRM distribution ($\sigma^2_{th}$) for a single-scale model of a randomly oriented field structure can be described by a simple model \citep{murgia2004}:
\begin{equation}\label{eq:rm_model}
    \sigma^{2_{RRM}}_{th} = 0.812^2 \left(\frac{\Lambda_c}{\text{pc}}\right) \int \left(\frac{n_e}{\text{cm}^{-3}} \frac{B_{||}}{\mu \text{G}}\right)^2 \frac{\text{d}l}{\text{pc}}, 
\end{equation}
where $\Lambda_c$ is the magnetic field reversal scale, $n_e$ is the gas electron number density, and $B_{||}$ is the magnetic field over any line-of-sight through the supercluster over a path length $L$.
The model $\sigma^{2_{RRM}}_{th}$ can be computed as Eq.\ref{eq:rm_model} for different values of the parameters $(\Lambda_c$, $B_{||}, L)$ and compared with the measured $\sigma^{2}_{MAD}(RRM)$ at the densities in the first two bins as derived in \Sec\ref{sec:resultrrm}.\par
We adopt a Bayesian Monte Carlo sampling (e.g. Monte Carlo Markov Chain, MCMC) approach to explore the likelihood surface and reconstruct the posterior distribution of the free parameters:  $(\Lambda_c$, $B_{||}, L)$. In the modeling framework, we call the data \textbf{d} and the model parameters \textbf{m}, so we can write the Bayes theorem (up to a constant) as: 
\begin{equation}\label{eq:bayestheorem}
P(\mathbf{m} \mid \mathbf{d)}\propto \mathcal{L}(\mathbf{d} \mid \mathbf{m}) \Pi, 
\end{equation}
that relates the posterior probability function of the parameters $P(\mathbf{m} \mid \mathbf{d)}$ to the likelihood function  $\mathcal{L}(\mathbf{d} \mid \mathbf{m})$, that we would like to maximize, and the prior $\Pi$, that encodes existing knowledge of parameter values.
In our case, we can write the likelihood function of detecting the observed $\sigma_{MAD}^{2_{RRM}}(n_e)$ for a medium of density $n_e$ as 
\begin{equation}
    \mathcal{L}\left(\sigma_{MAD}^{2_{RRM}} \mid \mathbf{m}\right) \propto \exp \left[-\frac{1}{2}\sum_i\left(\frac{\sigma^{2_{RRM}}_{MAD}(n_{e,i}) - \sigma_{th}^{2_{RRM}}(n_{e,i}, \mathbf{m})}{\delta(n_{e,i})}\right)^2\right],
\end{equation}
where $\delta$ is the uncertainty on the \textit{i}-th measured values of $\sigma_{MAD}^{RRM}$ and $\mathbf{m}=(\Lambda_c$, $B_{||}, L)$. For the parameters, we assume flat priors in ranges $0\leq B_{||}\leq 2$ $\mu$G, $5\leq\Lambda_{c}\leq 500 $ kpc, and $5\leq L\leq 70$ Mpc. While the choice of the parameters ranges is arbitrary given the limited information on supercluster environments, it is dictated by some reasonable considerations: the magnetic field strength is investigated up to values found in galaxy clusters (few $\mu$G, e.g. \citealt{bonafede2010, govoni2017}); the magnetic field will fluctuate on a range of scales, therefore we will investigate out to a large outer scale of $\Lambda_c=500$ kpc \citep{enblig2003, murgia2004, vacca2010}; finally, the maximum path length through the supercluster is computed following as the average of the three superclusters box size ($\sim70$ Mpc) defined in \cite{santiagobautista2020} that enclose each supercluster structure (see \T\ref{tab:mscc}). 

\begin{figure}[h!]
        \centering
        \includegraphics[width=1\linewidth]{ 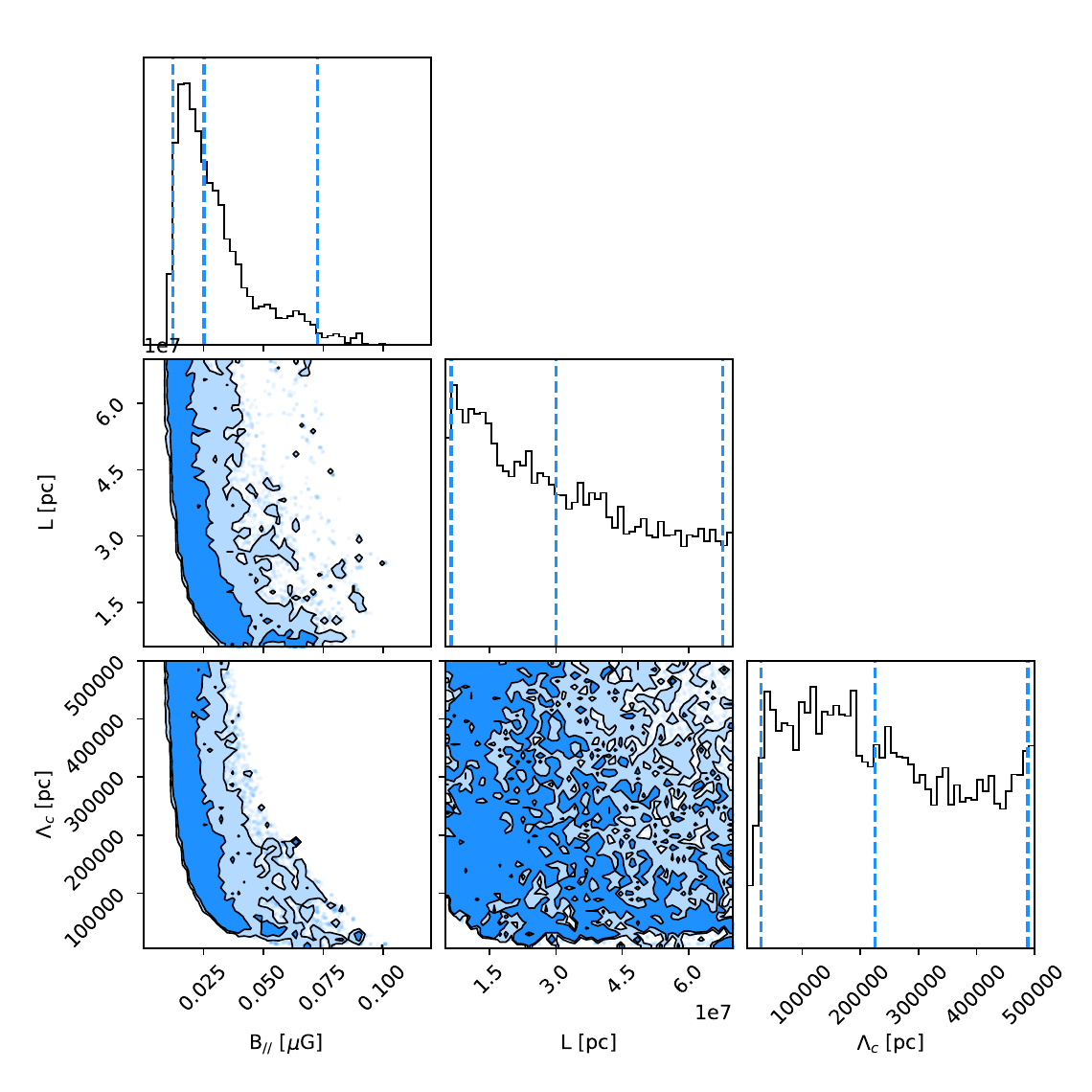}
        \caption[Corner plot]{Posterior probability distribution, marginalized into one and two dimensions, for the parameters $B_{||}, L,$ and $\Lambda_c$. Dark- and light-blue shaded areas indicate the 68 and 95\% confidence regions. The one-dimensional projection for each parameters is shown at the top. The dashed blue lines represent (from left to right) the 2.5th, 50th and 97.5th percentiles.}
        \label{fig:corner_plot}
    \end{figure}   
\begin{figure*}[h!]
    \centering
    \includegraphics[width=0.7\linewidth]{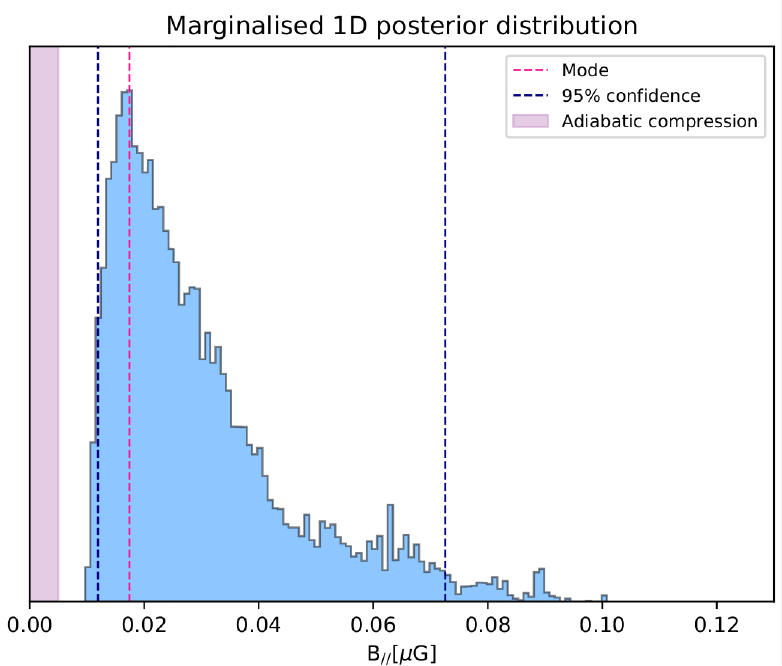}
    \caption[B-field marginalised posterior distribution]{Magnetic field marginalised posterior distribution, zoomed-in from \Fig\ref{fig:corner_plot}. The distribution is skewed towards small values. The 95\% confidence levels of the distribution are shown in dashed blue. The most likely value is shown in dashed red. The resulting adiabatic compression level is shown in purple. }
    \label{fig:Bfield}
\end{figure*}
The distribution is then sampled with MCMC using \texttt{emcee} \citep{emcee2013}. \Fig\ref{fig:corner_plot} shows the posterior probability distributions, marginalized into one and two dimensions, for the model parameters.
The distributions for each parameter are plotted along the diagonal and covariances between them under the diagonal. The shape of the covariance indicate the correlation between the parameters, namely a circular or diffuse covariance means no correlation, while elongated shapes can show correlation. As expected from \Eq\ref{eq:rm_model}, the path length L and the reversal scale $\Lambda_c$ are essentially unconstrained within the prior range. The posterior distribution of the magnetic field shows correlation with both other parameters but, once marginalized over them, shows a Gaussian-like behavior, though skewed towards high values. As shown in \Fig\ref{fig:Bfield}, we can constrain $B_{||}=19^{+50}_{-8}$ nG,  different from zero with a confidence level larger than 95\%. \par

This result for the magnetic field in low density regions of superclusters is in agreement with what is found with different methods investigating the filaments of the cosmic web \citep{carretti2022, carretti2023, vernstrom2023} and preliminary studies on superclusters with Faraday RMs \citep{xu2006, shishir2024}. Moreover,some early works on simulations have investigated the contribution of large scale structure filaments to the RM of background sources. These simulations suggest that the magnetic field intensity in filaments could range from 10 nG to 100 nG \citep{ryu2008, cho2009, akahori2010, vazza2015}, which still is in agreement with our findings. \par In our analysis we are considering the density contribution from supercluster members only, therefore neglecting the possible filament contribution and, in turn, overestimating the magnetic field intensity. While a better approach may be to compare directly with simulations of the RM variations from supercluster structures, we can estimate the effect of including an additional density component by increasing the density of the second bin (see \Fig\ref{fig:totalsigmaRRM}) with a 20\% excess from filaments, as recently measured with profiles of the SZ signal from stacked galaxy pairs \citep{degraaff2019}. The resulting magnetic field is consistent with the initial finding, and therefore still comparable with works of LSS filaments.  

Constraints on the magnetic field strength in cosmic structures are of key importance to investigate magnetogenesis scenarios. For this purpose, ad-hoc cosmological simulation of superclusters are needed but not available yet. However, we can compute a preliminary estimate of the predicted adiabatic compression-only effect on a primordial seed and compare with our results. For adiabatic compression, the magnetic field strength scales with density as $B=B_0(n/n_0)^{2/3}$.
With an initial seed of cosmological origin, as derived from the analysis of CMB anisotropies \citep[i.e. $B_0\sim 2$ nG,][]{planck2016} compression seems enough to explain the magnetic field strength derived in this analysis. However, recent LOFAR RRM measurements in IGM filaments suggest magnetic field seeds more than an order of magnitude below the limit derived with CMB \citep{neronov2024}.
Assuming an initial seed of $B_0\sim0.11$ nG \citep{carretti2023} at mean critical barionic density of $n_0\sim4\cdot10^{-31}$ g cm$^{-3}$, the adiabatic compression to the density $n\sim5\cdot10^{-29}$ gr cm$^{-3}$ as in our case, would yield $B\sim3.5$ nG, and $B_{||}=B/\sqrt{3} \sim 2$ nG. Comparing the adiabatic compression effect to the resulting $B_{||}$ from our analysis (see \Fig\ref{fig:Bfield}), it suggests that in superclusters of galaxies may be at play different mechanism such as dynamo amplification or AGN feedback, to further amplify the magnetic fields from $\sim2$ nG to the measured $\sim$20 nG. Another possible amplification mechanism is the presence of accretion strong shocks surrounding the filaments of the cosmic web, as observed in \cite{vernstrom2023}.\par
\begin{figure}
    \centering
    \includegraphics[width=1\linewidth]{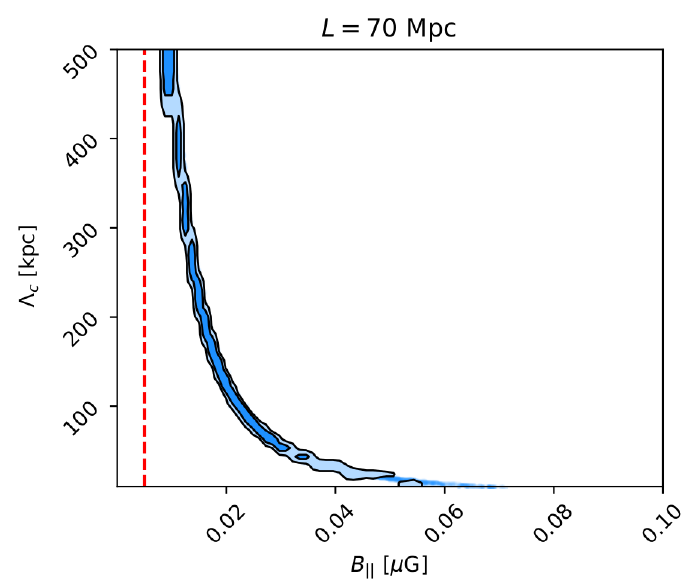}
    \caption{Covariance plot between the magnetic field strength ($B_{||}$) and reversal scale ($\Lambda_c$), fixing the path length to $L=70$ Mpc. The distribution 1- and 2-$\sigma$ contour levels are shown in black. The dashed red line represent the magnetic field resulting from adiabatic compression only. }
    \label{fig:Lmax}
\end{figure}

Since our result is dependent on the chosen $\Lambda_c$ parameter range, we can try to set an additional constrain on the reversal scale length by fixing the path length to $L=70$, the largest value considered for the parameter range. In this way, to reproduce the same RRM variance, we would minimize the magnetic field. The contour plot in \Fig\ref{fig:Lmax} shows the two dimensional posterior distribution between $B_{||}$ and $\Lambda_c$ once L is fixed to 70~Mpc. In the case of the maximum path length through the superclusters, the magnetic field is still larger than what is predicted by adiabatic-compression only for all values of $\Lambda_c$. A strong constrain on $\Lambda_c$ would result in the minimum line-of-sight magnetic field component in the supercluster environment.  

\section{Conclusions}\label{sec:conclusions}

The Faraday rotation measure signal from distant polarized sources is a sensitive probe of foreground, extragalactic low-density gas that permeates the large scale structure, otherwise difficult to detect. In this work, we have made use of the NVSS and LoTSS RM catalogues, combining the large number of sources found at cm-wavelengths with the highest precision of the sources detected at m-wavelengths, to investigate the rarefied environments of three major superclusters of galaxies in the northern sky: Corona Borealis, Hercules, and Leo.  Our findings can be summarized as follows:
\begin{itemize}
    \item For each supercluster we built a density 3D map (RA, DEC, z), computing the universal gas density profile \citep{pratt2022} for each member. 
    \item We created a catalogue of 4497 polarized sources fund in the background of the three superclusters. Among these, 3697 RM values are from the NVSS \citep{taylor2009}, and 818 are from LoTSS \citep{osullivan2023}. At the location of each source, we compute the mean gas density crossed by the polarized emission passing through the supercluster. We subtract the Galactic contribution, to investigate the extragalactic component. We then have for each source a pair of RRM and associated density values.
    \item We computed the MAD variance of the RRM distribution in three different density regimes: voids, filaments and nodes. While the highest density bin (inside galaxy cluster virial radii) is difficult to interpret due to several depolarization effects at play and to the low-number statistics, we detect an excess RRM variance of $\Delta\sigma^{2_{RRM}}_{MAD}=2.5\pm0.5$ rad$^2$ m$^{-4}$. between the lowest density region (outside the supercluster boundaries) and the low-density region inside the supercluster. We attribute this excess to the intervening medium of the filaments in the supercluster.
    \item We estimate the required magnetic field strength permeating the low-density environments crossed by the polarized emission through a Bayesian approach. We infer a line-of-sight magnetic field strength of $B_{||}=19^{+50}_{-8}$ nG, while other quantities such as the path-length and the magnetic field reversal scale remain unconstrained. Our result is in line with other work conducted on the filaments of the large scale structure \citep[e.g.][]{vernstrom2021, carretti2022, carretti2023}. Assuming only the effect of adiabatic compression from a primordial seed field consistent with current upper limits from LOFAR RRM measurements in IGM filaments \citep{neronov2024}, would give $B_{||}\sim 2$ nG. Our result suggests that in superclusters of galaxies different mechanisms of magnetic field amplification may be at play, such as dynamo amplification or AGN/galaxy feedback, or that an additional primordial seed should be considered.
\end{itemize}

\begin{acknowledgements}
SPO acknowledges support from the Comunidad de Madrid Atracción de Talento program via grant 2022-T1/TIC-23797, and grant PID2023-146372OB-I00 funded by MICIU/AEI/10.13039/501100011033 and by ERDF, EU. AB acknowledges financial support from the ERC Starting Grant `DRANOEL', number 714245. FV acknowledges the support by Fondazione Cariplo and Fondazione CDP, through grant n° Rif: 2022-2088 CUP J33C22004310003 for "BREAKTHRU" project.
\end{acknowledgements}

%
%
\bibliographystyle{aa}
\bibliography{bibliography}

\begin{appendix} 

\section{Inspection of FDF for source selection}\label{appendix}
\begin{figure*}[]
   \centering
   \includegraphics[width=1\linewidth]{ 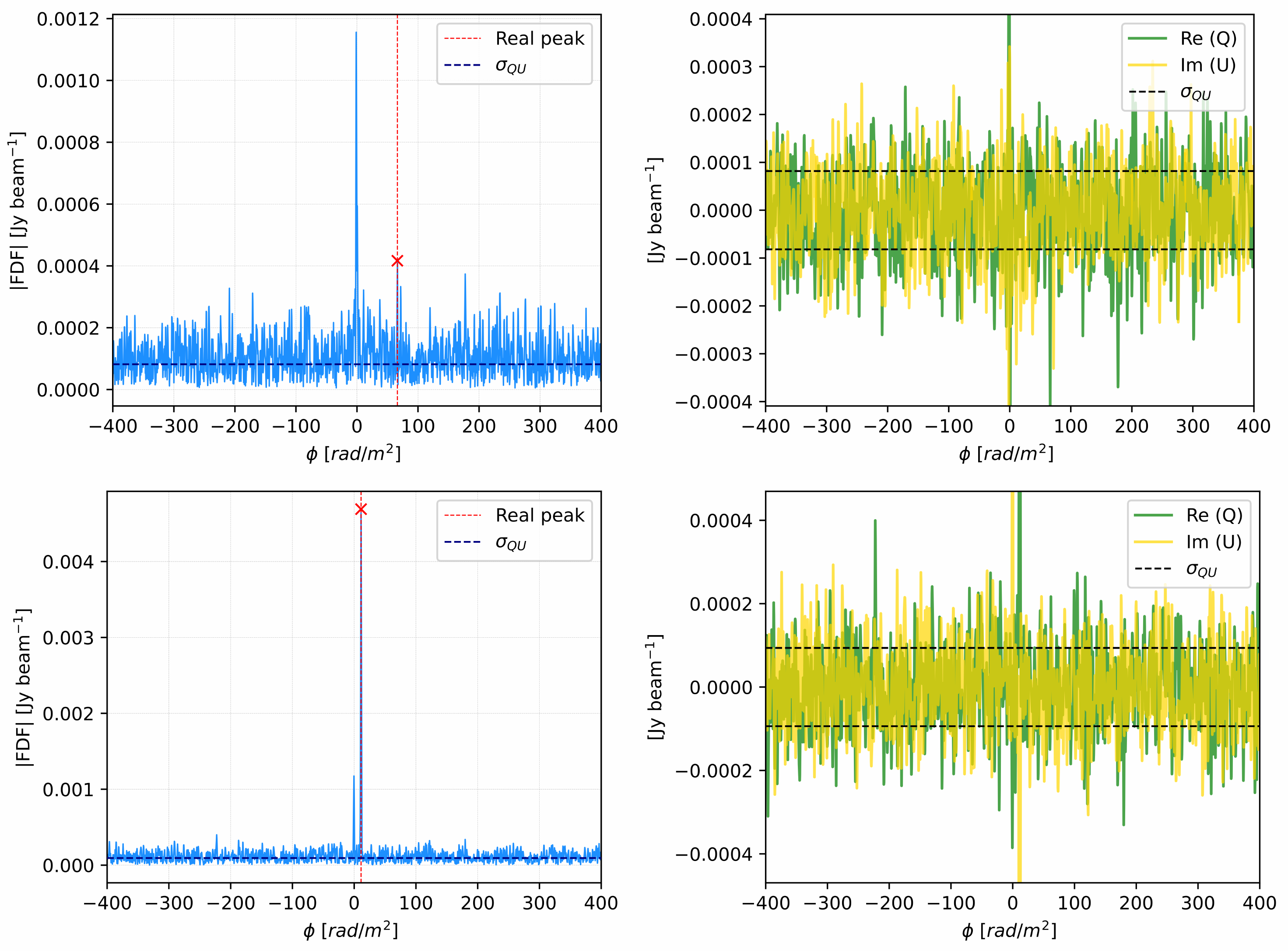}
      \caption[Faraday spectrum of polarized sources.]{We show the absolute value of the Faraday spectrum (FDF, blue solid line) and the real (Q, green solid line) and imaginary (U, yellow solid line) components of the FDF, for two example polarized source components. The Q,U plots have restricted range on the y-axis, to better visualize the noise. The 'real' peak, i.e. the highest signal-to-noise polarized component outside the leakage range, is marked with a red cross. The leakage peak is noticeable in both cases at $\phi\sim0$ rad m$^{-2}$. The rms noise ($\sigma_{QU}$) level is shown as a dashed line. After inspection, the top source at SNR$\sim8$ is excluded, while the bottom source, at SNR$\sim40$ is accepted.}
         \label{fig:plots_insp}
\end{figure*} 

\begin{figure*}
   \centering
   \includegraphics[height=0.9\textheight]{ 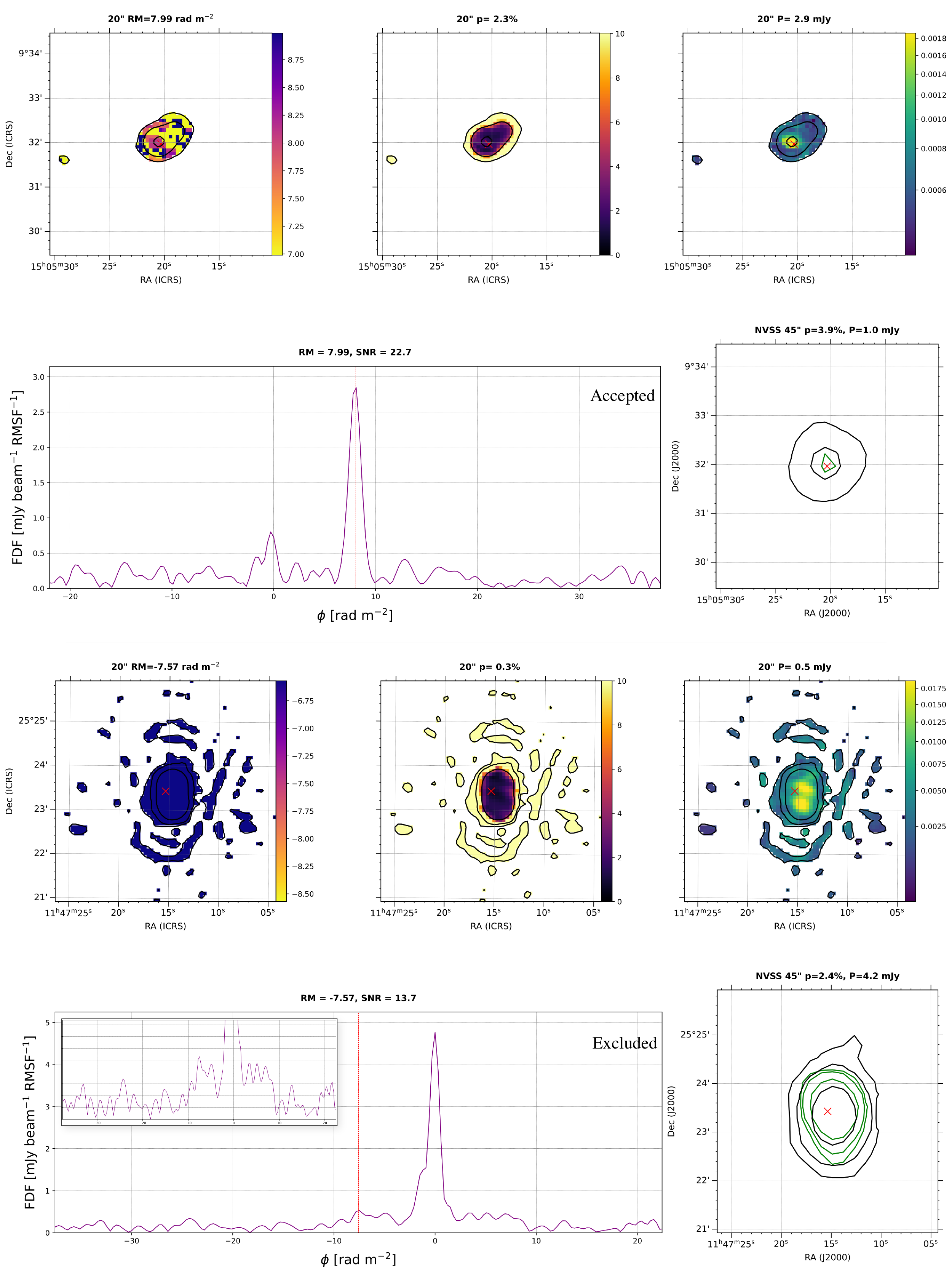}
   \caption[RM and polarized intensity inspection maps.]{Inspection maps to evaluate the catalogued polarized source component. \textit{Top row}: we show the LoTSS 20$''$ resolution RM map, degree of polarization map, and the polarized intensity map for the catalogued source inside the Stokes I contours (black). \textit{Bottom row}: we show the FDF with the catalogued RM value and SNR for the peak, and the NVSS 45$''$ resolution Stokes I (black) and polarized intensity (green) contours for comparison. For the bottom source, the zoomed-in panel show the complexity of the spectrum around the selected peak. After inspection, the top source is accepted, while the bottom source is excluded. }
         \label{fig:maps_insp}
\end{figure*} 

\end{appendix}
\end{document}